# Object Kinetic Monte Carlo Simulations of Radiation Damage in Bulk Tungsten Part-II: With a PKA Spectrum Corresponding to 14-MeV Neutrons

Giridhar Nandipati[a][1], Wahyu Setyawan[a], Howard L. Heinisch[a], Kenneth J. Roche[a,b], Richard J. Kurtz[a] and Brian D. Wirth[c]

[a] *Pacific Northwest National Laboratory, Richland, Washington, US*
[b] *Department of Physics, University of Washington, Seattle, WA 98195, USA*
[c] *University of Tennessee, Knoxville, Tennessee, USA*

**Abstract:** Object kinetic Monte Carlo was employed to study the effect of dose rate on the evolution of vacancy microstructure in polycrystalline tungsten under neutron bombardment. The evolution was followed up to 1.0 displacement per atom (dpa) with point defects generated in accordance with a primary knock-on atom (PKA) spectrum corresponding to 14-MeV neutrons. The present study includes the effect of grain size (2.0 and 4.0 μm) but excludes the impact of transmutation or pre-existing defects beyond grain boundary sinks. Vacancy cluster density increases with dose rate, while the density of vacancies decreases. Consequently, the average vacancy cluster size and the fraction of vacancies in visible clusters decrease with increasing dose rate. The density of vacancies and vacancy clusters decrease with grain size such that the average size of the clusters remains similar. However, the average size is larger for larger grains at dose rates < $4.5 \times 10^{-7}$ dpa/s. The trend of vacancy accumulation as a function of dose, dose rate, and grain size is similar to that obtained with the High Flux Isotope Reactor (HFIR) PKA spectrum. However, the amount of vacancy accumulation and the vacancy microstructure are quite different. Compared to the HFIR case, we find that even though the dose rates are 2.5 times higher, the density of vacancies and the average vacancy cluster sizes are lower. In addition, a void lattice forms only for the lowest two dose rates ($4.5 \times 10^{-8}$ and $4.5 \times 10^{-9}$ dpa/s). In contrast, a void lattice formed at all dose rates studied using the HFIR PKA spectrum. We discuss in detail the factors that lead to these different microstructures.

## 1. Introduction

At present, there are no known facilities that can irradiate materials or components with the necessary intensity and neutron energy spectrum needed to study material behavior under fusion-relevant conditions. Therefore, to explore radiation-induced degradation phenomena and microstructure evolution in W, a variety of neutron irradiation facilities are being used. Of these, only plasma-based devices provide a prototypic fusion neutron environment, but such facilities are currently much less attractive than alternative irradiation sources because of their high cost and low availability. [1,2] Most research fission reactors produce neutron fluxes significantly below those of envisioned fusion devices and also do not produce high-energy neutrons, which are characteristic of DT fusion. Correspondingly, the lower fluxes and less energetic neutrons in present research reactors are unable to produce damage levels beyond a few dpa/year. Note that the differences in neutron energy spectrum manifest as differences in the PKA spectrum produced by elastic collisions with neutrons. [3,4] The kinetics of microstructural evolution during irradiation is driven by the diffusion and reaction rates of various types of defects, and is significantly influenced by the size distribution of defect clusters created or produced during displacement cascades, which differ from one PKA spectrum to another. Furthermore, various neutron sources with different neutron spectra will produce different concentrations of both solid and gaseous transmutants in W. [5-8] Nevertheless, fission reactors and spallation neutron sources are being used to investigate neutron-induced damage in candidate fusion first-wall and plasma facing materials. Consequently, it is essential to construct physically-based quantitative models of microstructure evolution for use in analyzing and interpreting irradiation experiments performed using alternative irradiation sources. [9]

This article is Part-II of our four-part series of articles on object kinetic Monte Carlo (OKMC) simulations of damage accumulation in polycrystalline bulk tungsten due to neutron irradiation. Similar to our previous study [10], to isolate the effect of PKA spectrum, the effect of intragranular traps or pre-existing defects and transmutation on the defect accumulation is excluded. In this article, we present simulation results of damage accumulation in tungsten due to bombardment by 14-MeV neutrons. We explore doses up to 1 dpa with dose rates ranging from $4.5\times10^{-5}$ to $4.5\times10^{-9}$ dpa/s and grain sizes of 2.0 and 4.0 μm. In addition, we compare microstructural evolution due to 14-MeV neutron irradiation with those we previously modeled due to neutron irradiation conditions of HFIR [11,12]. Note that since the effects of intragranular traps and transmutation are not considered, the differences in the microstructural evolution are entirely due to differences in grain size and the production rates of defect clusters of various sizes between the two PKA spectra.

---

[1] Corresponding Author Email address: giridhar.nandipati@pnnl.gov (Giridhar Nandipati)
Tel.: +1(509) 375-2795, fax: +1(509) 375-3033

## 2. Simulation Details

The choice of the simulation cell dimensions and boundary conditions is described in detail in Part-I [9]. The list of kinetic parameters used in the present simulations are provided in Ref [10] and the description of the OKMC code, KSOME, is presented in Refs. [11, 12]. The only aspect that is different in this study from our previous study [9] is the PKA spectrum. As such, it will be discussed here. A detailed comparison of the PKA spectrum of HFIR and 14-MeV neutrons is presented below.

*2.1 Cascade Database*

An extensive database of primary defects from displacement cascades for $E_{MD}$ ($E_{PKA}$) ranging from 10-200 keV (12.5-276 keV) was generated using MD simulations. [13,14] Note the $E_{MD}$ denotes the amount of damage energy from the total kinetic energy of the PKA ($E_{PKA}$). [15,16] As an input for this OKMC simulation, each defect cluster is represented as an object. For the following discussion, a collection of these objects for a given $E_{MD}$ is simply referred to as a cascade. To insert the cascades into the simulation box according to the 14-MeV neutron PKA spectrum, we adopt the same procedure as described in Part-I [9], which is to create a new database of cascades of various $E_{MD}$ such that the proportion of cascades of a particular $E_{MD}$ is according to the normalized fraction of recoils at the corresponding $E_{PKA}$ in the PKA spectrum shown in Fig. 1. From Fig. 1, the average $E_{PKA}$ within the PKA range of 12.5-276 keV is approximately 100.3 keV, and the corresponding average $E_{MD}$ is approximately 75.0 keV. The average $E_{PKA}$ for the entire PKA spectrum is 44.78 keV. Dose rates for the full spectrum need to be scaled properly to yield the values for the 12.5-276 keV range before the corresponding cascade insertion rate is calculated. Note that all dose rates reported in this work are for the full-spectrum. With the above consideration, for dose rates between 4.5 x $10^{-5}$ and 4.5 x $10^{-9}$ dpa/s, the corresponding cascade production rates are 0.23814 x $10^{2}$ to 0.23814 x $10^{-2}$ cascades per second. Compared to the HFIR case, the cascade production rate is lower due to larger dpa produced per cascade at the respective average $E_{MD}$.

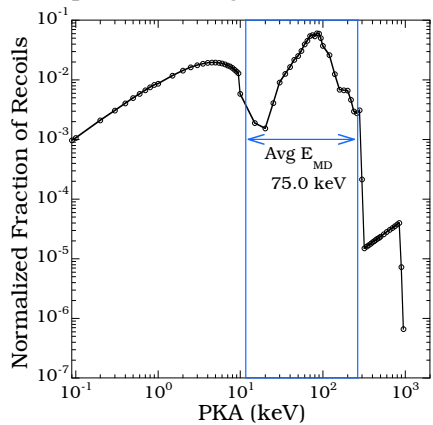

**Figure 1.** PKA spectrum for tungsten due to 14-MeV neutrons (NOTE: This PKA spectrum corresponds to a pure 14 MeV neutron source, not to a plasma device).

## 3. Results and Discussion

*3.1 General Details*

The results presented here are specifically for grain sizes of 2.0 and 4.0 μm at 1025 K, which is above the Stage III annealing temperature and thus (mono-) vacancies are mobile.[2] For all the dose rates studied, the density of vacancies and vacancy clusters decreases with increasing grain size. On the other hand, for both grain sizes, the vacancy cluster density increases with dose rate, but the number density of vacancies decreases. Consequently, the average size of vacancy clusters decreases with increasing dose rate. The damage accumulation (vacancy accumulation) trends with dose, dose rate and grain size are similar to those for tungsten neutron irradiated in accordance with the HFIR PKA spectrum. [10] Nevertheless, there are significant quantitative differences in the damage accumulation between the HFIR and 14-MeV cases. They are discussed in detail in Section 3.3.

---

[2] Note that in figures 5, 6 and 7, the factor of 4.5 factor is omitted, therefore, the value of $10^{-5}$ dpa/s represents a dose rate of 4.5 × $10^{-5}$ dpa/s.

*3.2 Microstructure of Neutron Irradiated W at 1.0 dpa*

Figures 2 and 3 show the vacancy cluster (void) microstructure at 1.0 dpa for dose rates from $4.5 \times 10^{-5}$ to $4.5 \times 10^{-9}$ dpa/s and grain sizes of 2.0 and 4.0 µm, respectively. The general characteristics of the microstructure as a function of dose and dose rate are similar for both grain sizes. The spatial ordering of voids along {011} planes is observed only for the lowest two dose rates of $4.5 \times 10^{-8}$ dpa/s and $4.5 \times 10^{-9}$ dpa/s, and the interplanar void spacing increases with grain size. For the dose rate of $4.5 \times 10^{-9}$ dpa/s, the spatial ordering of voids along {011} planes is clearly seen when viewed from any <100> direction. However, at the higher dose rates, the spatial distribution of vacancy clusters appears to be random. In contrast, spatial ordering of vacancy clusters was observed for all dose rates studied in the HFIR irradiation case, namely $1.7 \times 10^{-5}$– $1.7 \times 10^{-9}$ dpa/s. The average interplanar spacings for dose rates of $4.5 \times 10^{-8}$ and $4.5 \times 10^{-7}$ dpa/s are approximately 17.09 and 22.79 nm for 2.0 and 4.0 µm grains, respectively. These spacings are similar to those obtained in the HFIR study. [9] The differences in void microstructure are due to differences in the production rates of defect clusters of various sizes between the neutron irradiation with HFIR and 14-MeV PKA spectra. As expected, differences in damage accumulation and the associated microstructure evolution in tungsten strongly depend on the PKA spectrum of the radiation source. Even though no explicit dependence of the void lattice spacing on the dose or dose rate or PKA spectrum was observed, the appearance of the spatial ordering of vacancy clusters at a particular dose, i.e. the critical dose, seems to depend on both the dose rate and PKA spectrum of the radiation source, as well as the grain size of tungsten. Note that the aim of this work is to study the effect of PKA spectra on the irradiation damage in tungsten and not to explore the conditions or requirements to observe the void lattice formation. Nevertheless, we discuss later in this article about the conditions that are required to see the formation of a void lattice, solely based on our present KMC simulations.

*3.3 Damage Accumulation*

Figure 4 shows the density of vacancies and vacancy clusters, the average vacancy cluster diameter[3] and the fraction of vacancies in visible clusters, as a function of dose, dose rate, and grain size. With increasing grain size, self-interstitial atom (SIA) clusters have to travel farther before they reach a grain boundary, thereby increasing the probability of recombination with vacancies. Accordingly, the density of both vacancies and vacancy clusters decrease with grain size. With increasing grain size, the fraction of vacancies in visible clusters and average vacancy cluster size increase for the dose rates of $4.5 \times 10^{-8}$ and $4.5 \times 10^{-9}$ dpa/s (lowest) and for the higher dose rates the change is negligible. Figures 4(a-d) show that the density of vacancies decreases while the density of vacancy clusters increases with dose rate for both grain sizes. Consequently, the average vacancy cluster size decreases with increasing dose rate (Figs. 4(e-f)). The qualitative behavior of damage accumulation with dose, dose rate and grain size is similar to that observed in the simulations of microstructural evolution in neutron irradiated tungsten in HFIR. [10] However, in contrast to the HFIR case, the density of vacancies and the average vacancy clusters size saturate for higher dose rates ($\geq 4.5 \times 10^{-7}$ dpa/s), but increases with dose for the lower dose rates. Saturation of vacancy densities and average vacancy cluster size confirms that the spatial distribution of vacancy clusters at 1.0 dpa is random. Similar to the HFIR case, the appearance of the void lattice is delayed to a higher critical dose with increasing grain size even for the case of 14-MeV PKA spectrum.

Figures 4(g-h) show the fraction of vacancies that are part of visible clusters as a function of dose, dose rate and grain size. The fraction of vacancies in visible clusters increases with grain size for the dose rate of $4.5 \times 10^{-9}$ dpa/s, but decreases for the dose rate of $4.5 \times 10^{-8}$ dpa/s. The fraction of vacancies in visible vacancy clusters at 1.0 dpa for the dose rate of $4.5 \times 10^{-9}$ dpa/s is about 78% and 90% for 2.0 and 4.0 µm grains, respectively. However, for the dose rates of 4.5 x

---

[3] Vacancy clusters of all sizes were assumed to be spherical.

$10^{-7}$ dpa/s and higher, the fraction of vacancies in visible clusters is insignificant for both grain sizes. Similar to the HFIR irradiations, the onset of visible cluster[4] appearance shifts to higher doses with increasing dose rate and grain size.

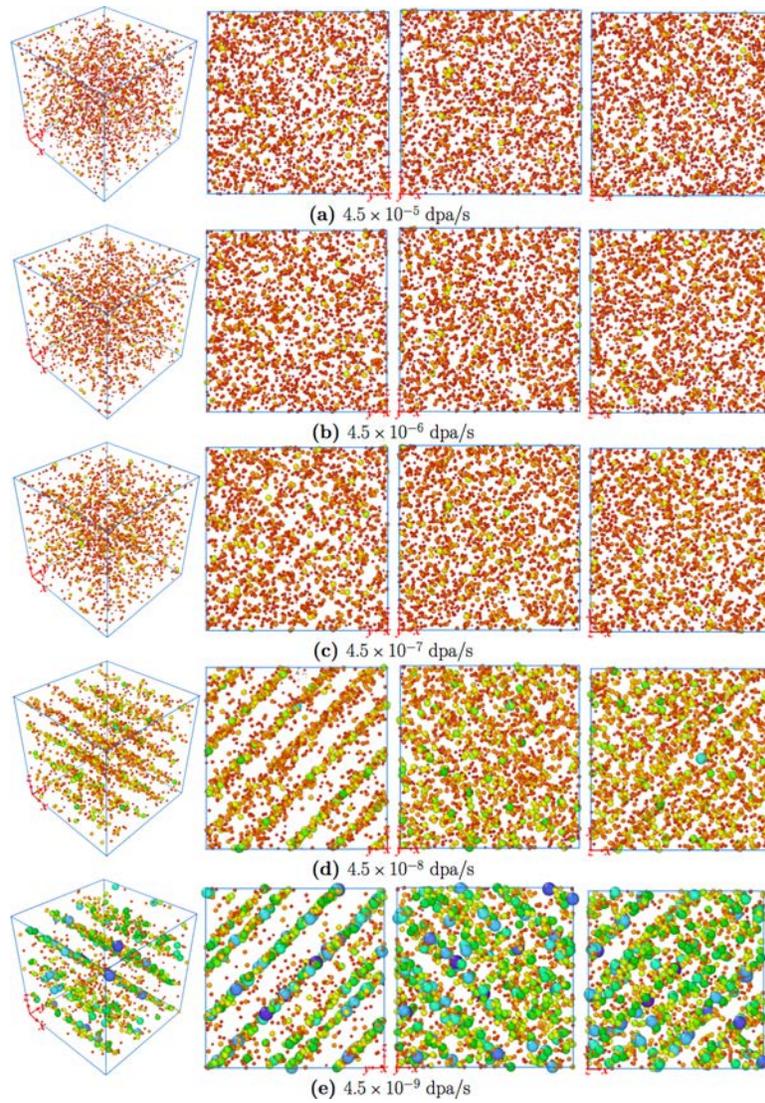

**Figure 2. Vacancy cluster microstructure at 1.0 dpa for 2.0 μm grain size. From left to right, when viewed in perspective and along [100], [010] and [001].**

---

[4] A diameter larger than 2.0 nm (approx. 300 vacancies) is considered visible under Transmission Electron Microscope examination.

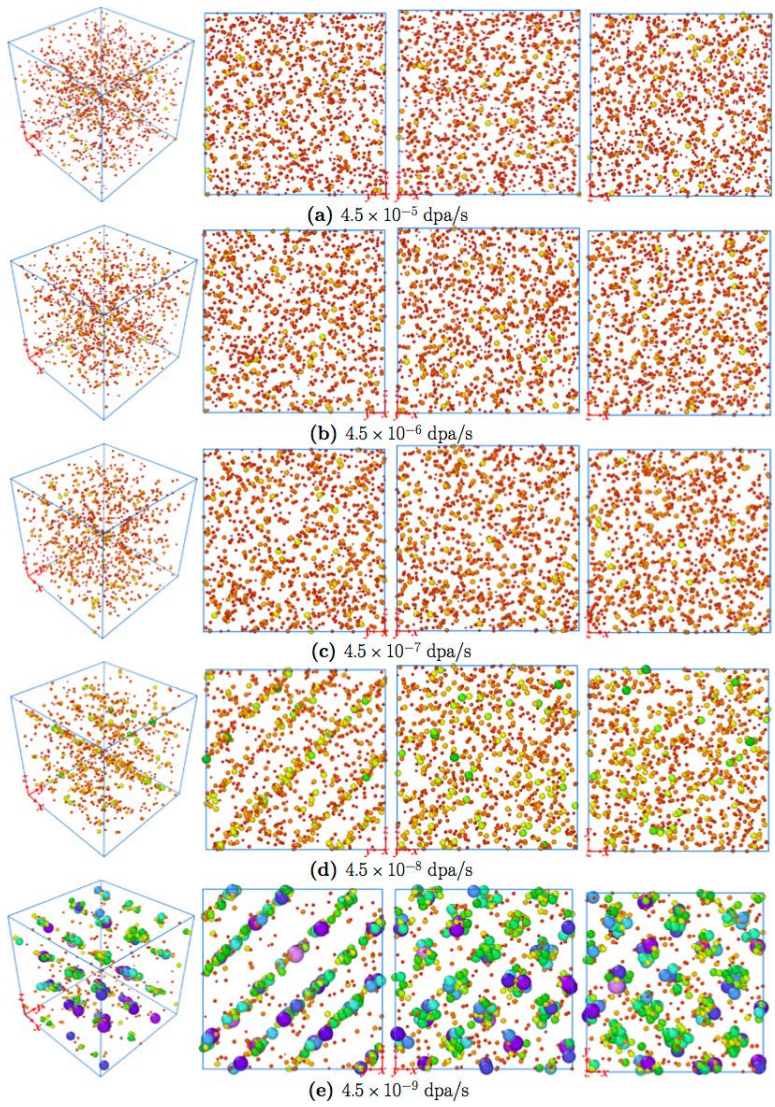

**Figure 3.** Vacancy cluster microstructure at 1.0 dpa for 4.0 μm grain size. From left to right, when viewed in perspective and along [100], [010] and [001].

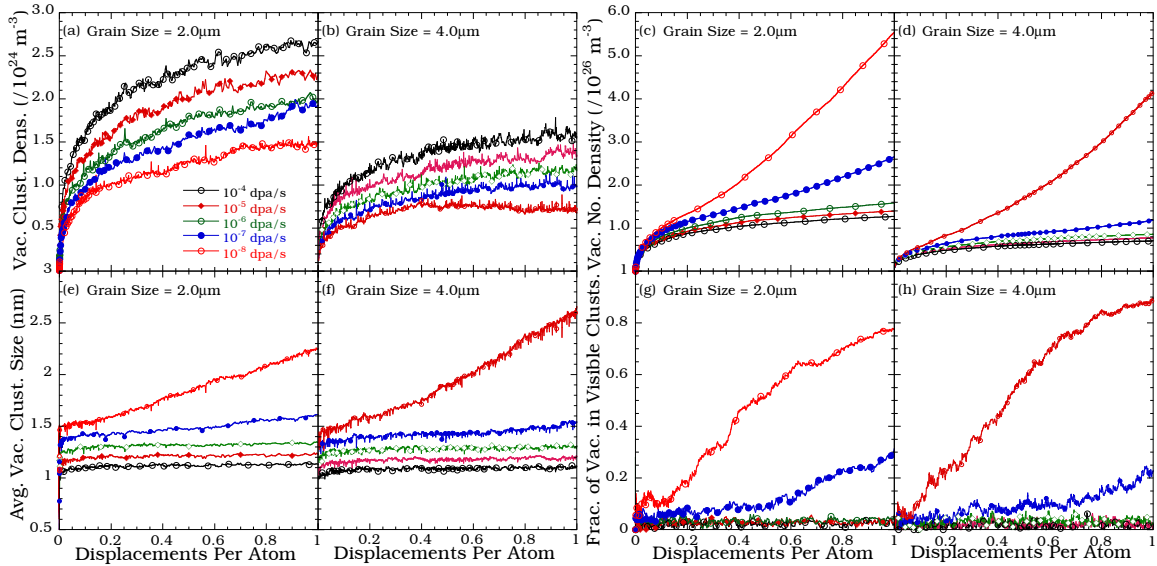

**Figure 4.** (a-b) Density of vacancy clusters, (c-d) density of vacancies, (e-f) average diameter of vacancy clusters, and (g-h) fraction of vacancies in the visible clusters (sizes larger than 2 nm), as a function of dose for various dose rates.

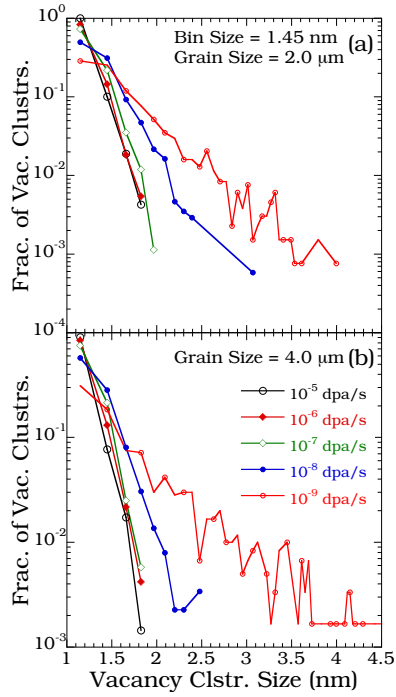

**Figure 5.** Binned vacancy cluster size distribution at 1.0 dpa, for grain sizes of (a) 2.0 μm and (b) 4.0 μm.

Figures 5(a-b) show the vacancy cluster size distribution with a bin size of 100 vacancies or a diameter of 1.45 nm, at a dose of 1.0 dpa. It clearly shows that the vacancy cluster size distribution becomes wider with decreasing dose rate, i.e. larger clusters are found at lower dose rates. For the dose rate of $4.5 \times 10^{-5}$ dpa/s, a large fraction of the vacancy clusters is smaller than 300 vacancies and the size distribution is narrow. In general, the behavior is similar, but the distributions in the present study are narrower compared to those observed in HFIR simulations. Also, for dose rates greater than or equal to $4.5 \times 10^{-7}$ dpa/s, the vacancy cluster size distribution appears to be similar for all dose rates and both grain sizes.

*3.3 Comparison of Damage Accumulation due to Neutron Irradiation with 14-MeV Neutron and HFIR PKA Spectra*

*3.3.1 Comparison of Defect Production*

As noted earlier, since the effects of intragranular traps and transmutation are not considered here, differences in damage accumulation are entirely due to differences in the production rates of defect clusters of various sizes. Therefore, we will first discuss how the defect cluster size distribution varies with cascade damage energy and differences in the corresponding defect production rates for the two PKA spectra. Note that in this discussion, any defect cluster of size less than or equal to 5 is considered to be a small cluster.

| Defect Type | $E_{MD}$ (keV) | | | | | | | | | |
|---|---|---|---|---|---|---|---|---|---|---|
| | 10 | 20 | 30 | 40 | 50 | 60 | 75 | 100 | 150 | 200 |
| SIA (> 5) | 0* (0**) | 3 (20) | 9 (29) | 21 (40) | 22 (45) | 39 (60) | 45 (65) | 52 (69) | 57 (-) | 61 (-) |
| Vac. (≥ 5) | 0 | 2 | 0 | 5 | 13 | 11 | 31 | 34 | 48 | 49 |
| Vac. (≥ 10) | 0 | 0 | 0 | 2 | 2 | 8 | 28 | 31 | 45 | 46 |

*Before cascade annealing, as obtained from MD simulations    ** After 1.0 ns of cascade annealing [17]

**Table 1.** Fraction (%) of SIAs in clusters greater than size 5 and the fraction of vacancies in clusters larger than size 5 and 10 produced in tungsten cascades of various $E_{MD}$ energies

Table 1 shows the in-cascade defect clustering as a function of cascade damage energy ($E_{MD}$). For SIAs, Table 1 also shows the in-cascade clustering obtained from MD simulation at a time of ~ 55 ps, and after a 1.0 ns cascade anneal at 1025K [17] in parentheses. SIA clustering after cascade annealing is always larger due to capture of individual SIAs by larger SIA clusters and it is the effective SIA clustering observed outside the cascade volume. In-cascade clustering of both SIAs and vacancies increases with damage energy. However, the increase in clustering is significant only for cas-cades with $E_{MD} \geq 50$ keV. In addition, SIA clusters larger than size 5 can only diffuse in 1D and an increase in their fraction reduces cascade annealing due to intracascade recombination. Moreover, any vacancy clusters smaller than size 5 will dissociate at 1025 K, mostly via vacancy emission. Accordingly, one can say that the effective vacancy clustering in cascades of $E_{MD} \leq 30$ keV is almost zero.

Between 10 keV $\leq E_{MD} \leq 200$ keV, the production rates of cascades of various $E_{MD}$ are quite different between the two PKA spectra. For the 14-MeV PKA spectrum, approximately, 83% of the cascades produced have $E_{MD} \geq 50$ keV, of which 43% (26%) of them have $E_{MD} \geq 75$ keV (100 keV). While for the HFIR PKA spectrum, 95% of the cascades produced have $E_{MD} < 50$ keV, of which 66% (52%) have $E_{MD} \leq 20$ keV (10 keV). Accordingly, the average $E_{MD}$ for the HFIR and 14 MeV PKA spectra over the range of 10 to 200 keV is approximately 21 keV and 75 keV, respectively. For a given dose rate, compared to the HFIR case, the cascade production rate is lower in the case of 14-MeV irradiation due to the larger dpa per cascade at their respective average $E_{MD}$.

The production rate of mono-vacancies and small vacancy clusters is higher than that of large vacancy clusters for both PKA spectra. However, the vacancy and small vacancy cluster production rate is much higher for HFIR irradiation. The production rate of large SIA clusters is higher under 14-MeV irradiation while that of small SIA clusters is higher in HFIR. Moreover, for 14-MeV irradiation, on average only one cascade is necessary to reach a damage level equivalent to a 75 keV cascade, while for HFIR irradiation, approximately four cascades of 21 keV are needed, which are separated both in time and space, to reach a similar damage level. Therefore, damage accumulation occurs more uniformly in HFIR than under 14-MeV irradiation.

*3.3.2 Comparison of Microstructural Evolution*

Figure 6 shows ratios between vacancy densities, vacancy cluster densities, and vacancy cluster sizes of 14 MeV and that of HFIR PKA spectra (14-MeV/HFIR) at various dose rates. Note that the ratios are for dose rates with same exponent (e.g. 4.5 x $10^{-5}$ dpa/s for 14 MeV and 1.7 x $10^{-5}$ dpa/s for HFIR). Because the neutron spectrum of 14 MeV is harder than HFIR, one would expect that vacancy accumulation at 1.0 dpa should be higher than in HFIR. Contrary to this expectation, Figs. 6(a, a') show that at 1.0 dpa, the vacancy density ratios are less than one for both grain sizes for all dose rates, except for the dose rate of $10^{-5}$ dpa/s for 4.0 μm grain size. This result is particularly surprising because dose rates for the 14-MeV PKA spectrum are 2.5 times higher than for the HIFR PKA spectrum. With an exception for 4.0μm grain size, values of the vacancy cluster density ratios at 1.0 dpa are greater than one except for the higher dose rates (Figs. 6(b, b')). Accordingly, the average size of vacancy clusters is also smaller in the case of 14-MeV PKA spectrum (Figs. 6(c, c')) with the exception for the highest dose rate for 4.0 μm grain size. Nevertheless, it is important to note that the above comparison of accumulated damage is made at a similar dose (or fluence). This does not necessarily mean that similar dose levels are reached at the same real time. The damage accumulation rate as a function of time is faster for the 14-MeV PKA spectrum.

Under 14-MeV irradiation, the production rates of both large vacancy and SIA clusters are higher than in HFIR. During the initial stages of irradiation, because of significant SIA clustering, most of the large SIA clusters diffuse to a grain boundary intact (minimal contribution to intercascade recombination). While large vacancy clusters take longer to dissociate. Therefore, as expected, during the initial stages of irradiation, damage accumulates faster with increasing dose for the 14-MeV PKA spectrum. However, as the dose increases, the probability for an SIA cluster to recombine with a vacancy or vacancy cluster increases and damage accumulation slows down and eventually saturates when the probability approaches one (all newly created SIAs recombine and do not reach a grain boundary). Except for the highest dose rates of 4.5 x $10^{-5}$ and 4.5 x $10^{-6}$ dpa/s, the slowing down of damage accumulation occurs before 0.2 dpa. Nevertheless, from Figs. 6(a, a'), it can be seen that at 1.0 dpa the vacancy density ratio decreases with increasing dose and drops below one except for the for the higher dose rate for 4.0 μm grain size.

For all dose rates studied in the case of 14-MeV irradiation, the vacancy cluster density ratio is greater than one at 0.2 dpa (Figs. 6(b, b')) for both grain sizes due to higher production rates of large vacancy clusters. However, with increasing dose, this ratio decreases with increasing dose for both grain sizes. For the highest dose rates ($\geq 4.5$ x $10^{-7}$ dpa/s) and, for $10^{-9}$ dpa/s in the case of 4.0 μm grain size, the ratio either drops below one or is very close to one. Although for lower dose rates (4.5 x 10$^{-8}$ and 4.5 x 10$^{-9}$ dpa/s) the ratio decreases with increasing dose, it still stays higher than one even at 1.0 dpa. We think that for HFIR irradiation, the vacancy cluster density decreases considerably at the lower dose rates (1.7 x 10$^{-8}$ and 1.7 x 10$^{-9}$ dpa/s) because of Ostwald ripening or coarsening during the time interval between cascades, which is longer than in the case of 14-MeV irradiation. For the dose rate of 10$^{-9}$ dpa/s and 4.0 μm grain size, the ratio drops below one due to the saturation of vacancy cluster density in the case of 14-MeV irradiation. Note for the lowest dose rate for both grain sizes the ratio of average cluster sizes first decreases and starts to increase with increasing dose and is more obvious in the case of 4.0 μm grain size.

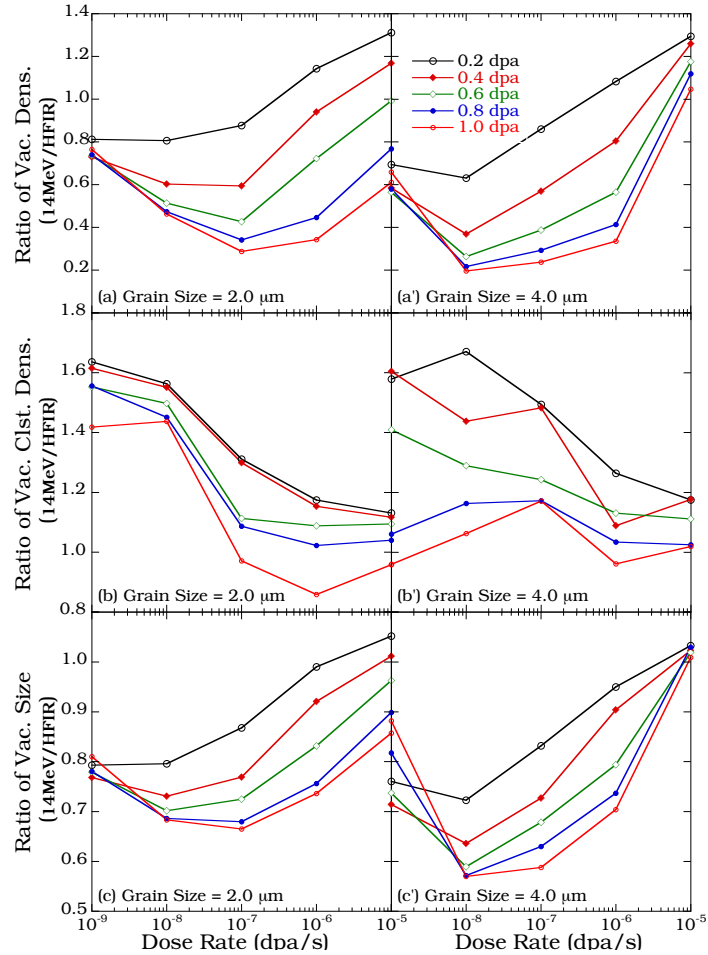

**Figure 6. Ratio of (a, a') the density of vacancies, (b, b') the density of vacancy clusters, and (c, c') the average vacancy cluster size at various doses and dose rates for grain sizes of 2.0 and 4.0 μm, respectively, between 14-MeV neutron and HFIR PKA spectra. Ratios are for the dose rates with same exponent.**

From Table 1 it can be seen that in-cascade vacancy clustering is minimal for cascades of $E_{MD}$ < 50 keV. However, it is surprising to see larger vacancy clusters in the HFIR case than for the 14-MeV case. Note that for the lowest dose rate, vacancy clusters as large as size 15 will dissociate before the next cascade occurs. A back-of-the-envelope[5] calculation shows that, no matter how small the production rate is, a sizeable number of cascades with $E_{MD} \geq 50$ keV are produced. A fraction of large vacancy clusters produced by high-energy cascades survive and subsequently grow due to a steady supply of mono-vacancies produced from low-energy cascades or from the dissociation of small vacancy clusters. Therefore, for the HFIR PKA spectrum, during the initial stages of irradiation, both the vacancy and vacancy cluster densities increase at a slower rate with dose. However, due to a continuous supply of mono-vacancies, the vacancy density continues to grow slowly until the formation of a void lattice, after which, it increases rapidly.

---

[5] $v_{NRT}^{21keV}$ = 65.6 displacements, number of cascades inserted into the simulation cell to reach 1.0 dpa (approx.) = 850500. Number of cascades inserted in the simulation cell with $E_{MD} \geq 50$ keV = 42525 cascades.

*3.3.3 Void Lattice Formation*

Another big difference in the microstructure between the two PKA spectra is the appearance of a void lattice at all dose rates studied for the HFIR PKA spectrum while only at the lowest two dose rates studied for the 14 MeV PKA spectrum. Although no relationship between the PKA spectrum, dose or dose rate on the void lattice interplanar spacing can be made, these results show that the production rates of defect clusters of various sizes, which in turn depend on the PKA spectrum and dose rates of a radiation source, have a significant effect on the appearance or absence of a void lattice at low-dose in tungsten

The qualitative behavior of faster damage accumulation with dose during the initial stages of irradiation and the subsequent slowing down of damage accumulation due to increased intercascade recombination is common to both PKA spectra. Yet, the quantitative damage accumulation depends on the production rates of defect clusters of various sizes and the time interval between two consecutive cascade insertions. Within this time interval, the following processes occur chronologically; (1) large SIA clusters either recombine with a vacancy or diffuse to a grain boundary, while at the same time small SIA clusters (size≤5) either recombine or are captured by larger SIA clusters, these processes occur on a nanosecond to millisecond time scale, (2) diffusion and capture of freely migrating mono-vacancies[6] (FMV) by larger vacancy clusters, which occurs on a millisecond time scale, (3) dissociation of vacancy clusters and capture of emitted mono-vacancies by other vacancy clusters. Depending on the size of vacancy clusters, these processes occur over a time scale of milliseconds to minutes. Note that the time interval between cascade insertions ranges from milliseconds[7] for the highest dose rates to minutes[8] for the lowest dose rates.

For higher dose rates, cascades are inserted on a time scale similar to that of the diffusion and capture of FMVs by vacancy clusters from a different cascade and the dissociation of small vacancy clusters. This increases the probability of FMVs being annihilated by a 1D diffusing SIA cluster, resulting in a lower accumulation of vacancies and a smaller average vacancy cluster size with increasing dose. Moreover, an increase in the probability for a 1D diffusing SIA cluster to annihilate with FMVs results in a slow attrition of its size and when it is reduced to size 5 or lower, it will perform 3D diffusion. In addition, due to the faster cascade production rate, the vacancy cluster density grows more rapidly with dose. The overall effect of a reduction in the fraction of SIA clusters that can diffuse in 1D and faster production of vacancy clusters is a delay in the formation of a void lattice, that is, its formation is pushed to a higher dose or even prevented. To check this assertion, we have carried out an additional set of simulations using the HFIR PKA spectrum up to 1.0 dpa and with SIA clusters up to size 10 allowed to rotate, which effectively reduces the fraction of 1D migrating SIA clusters contributing to the formation of the void lattice. We observed no void lattice for the highest dose rates. [18]

For lower dose rates, the time interval between cascades insertions is on a time scale of seconds to minutes. The time interval is long enough that almost all the FMVs are absorbed by vacancy clusters well before the next cascade occurs, resulting in an increase in both the average size and the number density of vacancies. In addition, vacancy cluster density decreases due to the dissociation of small vacancy clusters that are created during a cascade. The overall effect is an increase in the effectiveness of 1D diffusing SIA clusters either to partially or fully annihilate unfavorably positioned vacancy clusters. Partially annihilated vacancy cluster(s) will dissociate and the released mono-vacancies are captured by favorably positioned larger vacancy clusters. This not only reduces the density of vacancy clusters further but also aids the formation of a void lattice. Note that the dissociation of partially annihilated vacancy clusters on their own and the capture of released mono-vacancies by other vacancy clusters is essential for the formation of the void lattice, which is why the void lattice did not form in our irradiation simulations at 300 K [19].

For the 14-MeV PKA spectrum, both the production of large vacancy and SIA clusters is higher compared to the HFIR PKA spectrum. Moreover, because of higher in-cascade SIA clustering, cascade annealing due to intracascade recombination is lower. Higher SIA clustering not only reduces intracascade recombination but also intercascade recombination during the initial stages of irradiation. Therefore, during the initial stages of irradiation, damage accumulation occurs at a

---

[6] Mono-vacancies that survive intra-cascade recombination and become spatially uncorrelated with original cascade.
[7] 42 ms and 12 ms for 14 MeV and HFIR PKA spectra, respectively.
[8] 420 s and 120 s for 14 MeV and HFIR PKA spectra, respectively.

faster rate as a function of dose. We think that the absence of a void lattice at higher doses is caused by a significantly higher vacancy cluster density coupled with a larger average vacancy cluster size during the initial stages of irradiation. As noted earlier, dissociation of a partially annihilated vacancy cluster is essential for the formation of a void lattice. However, larger partially-annihilated vacancy clusters would require more time to dissociate. Accordingly, at higher dose rates, vacancy clusters are produced more frequently than they are annihilated by 1D diffusing SIA clusters. While at lower dose rates, the time interval between cascade insertions is long enough to allow those clusters to dissociate. Consequently, a void lattice is observed at lower dose rates while a random spatial distribution of vacancy clusters is observed at higher dose rates. To test this assertion, we extended the irradiation simulation for the highest dose rate (4.5 x $10^{-6}$ dpa/s) up to 2.0 dpa. As expected, the spatial distribution of vacancy clusters remains random even at 2.0 dpa. In addition, to test the effect of vacancy cluster size on the formation of the void lattice, we have carried out two sets of ad-hoc irradiation simulations, one with only 20-keV cascades and the other only with 100-keV cascades for dose rates from $10^{-4}$ to $10^{-7}$ dpa/s and for a 2.0 µm grain size. We find that void lattice formation was observed only for the highest dose rate studied in the case of 20-keV cascades and only for the lowest dose rate studied in the case of 100-keV cascades (see Fig. 7). For the irradiation simulation using only 20-keV cascades, higher dose rates are required to promote the formation of vacancy clusters that are large enough to avoid dissociation. While for the irradiation simulations using only 100-keV cascades, lower dose rates are required so that there is enough time for partially annihilated vacancy clusters to dissociate, thus aiding formation of the void lattice. This behavior of void lattice formation is in agreement with one the proposed conditions to observe void lattice formation by Ghoniem *et al.*[20], which is the formation immobile vacancy clusters during the cascade formation. More importantly, the interplanar spacing is smaller in the case of 20-keV cascades in which the inserted cascades have minimal vacancy clustering compared to the case of 100-keV cascades in which the inserted cascades have more vacancy clustering. These results suggest that the interplanar spacing is influenced by the in-cascade vacancy clustering

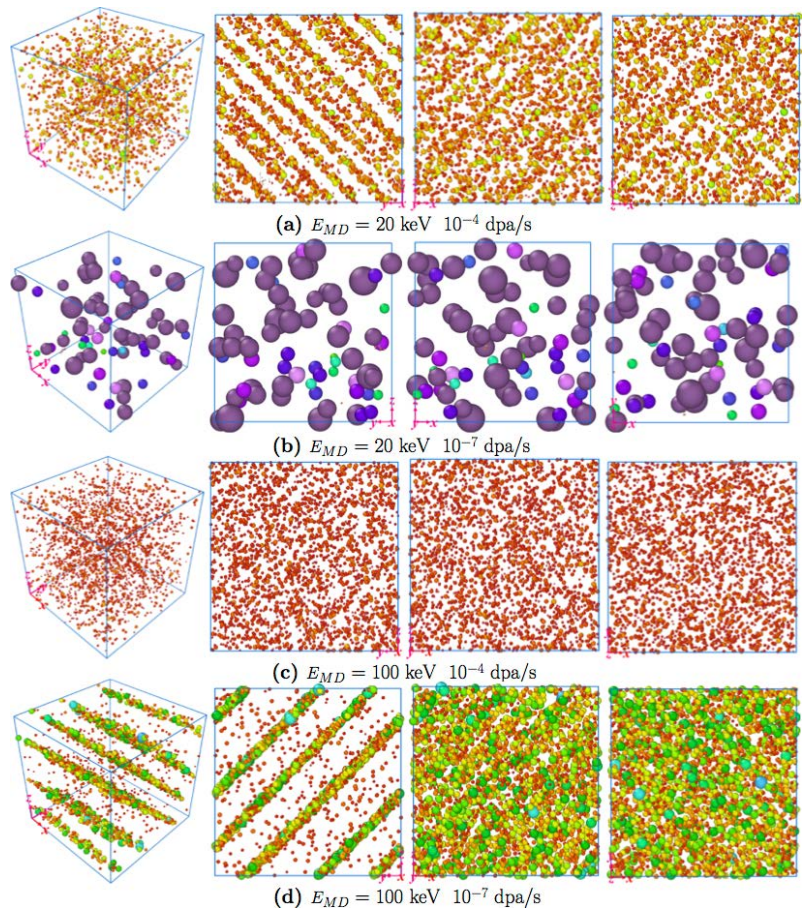

(a) $E_{MD} = 20$ keV $10^{-4}$ dpa/s

(b) $E_{MD} = 20$ keV $10^{-7}$ dpa/s

(c) $E_{MD} = 100$ keV $10^{-4}$ dpa/s

(d) $E_{MD} = 100$ keV $10^{-7}$ dpa/s

**Figure 7. Vacancy microstructure at 1.0 dpa for 2.0 µm grain size when viewed in perspective and along [100], [010] and [001] directions.**

## 4. Conclusions

OKMC simulations were carried out to examine damage accumulation in polycrystalline bulk tungsten irradiated at various dose rates and grain sizes due to 14-MeV neutrons at a temperature of 1025 K. The qualitative behavior of damage accumulation with dose, dose rate, and grain size is similar to HFIR irradiation simulations. That is, for a given dose, with increasing dose rate: 1) the density of vacancies decreases, 2) the average size of vacancy clusters decreases, and 3) the density of vacancy clusters increases. However, quantitatively the damage accumulation caused by the two PKA spectra is quite different. For the 14-MeV PKA spectrum, even though the damage accumulates at a faster rate with increasing dose, it saturates for higher dose rates, while for the lowest two dose rates damage continues to accumulate because of the formation of a void lattice.

Vacancy clusters form a void lattice only for the lowest two dose rates while their distribution is random at higher dose rates. This is in contrast to the HFIR case where a void lattice forms at all dose rates. The critical dose for a void lattice to form appears to depend on the PKA spectrum, the dose rate of the radiation source, and the grain size. The critical dose shifts to a higher value for larger grains and dose rates. More importantly, the PKA spectrum influences void lattice formation such that, depending on the in-cascade vacancy clustering, a dose rate which is too high or too low inhibits the formation of a void lattice. In addition to the diffusion of mono-vacancies and the dissociation of small vacancy clusters, the dissociation of partially annihilated vacancy clusters is also necessary for the appearance of a void lattice. Ad-hoc irradiation simulations with cascades of $E_{MD}$ = 20 and 100 keV suggest that the interplanar spacing of the void lattice depends on in-cascade vacancy clustering. In addition, Fig. 7 shows that one cannot simulate the radiation damage by simply using cascades whose damage energy is equal to the average of a PKA spectrum.

## ACKNOWLEDGEMENT

The work described in this article was performed at Pacific Northwest National Laboratory (PNNL), which is operated by Battelle for the United States Department of Energy (US DOE) under Contract DE-AC06-76RL0-1830. The US DOE, Office of Fusion Energy Sciences (FES) and Office of Advanced Scientific Computing Research (ASCR) supported this study through the SciDAC-3 program. All computations were performed on HOPPER at the National Energy Research Scientific Computing Center (NERSC) and PNNL's Institutional Resources (PIC). The authors would like to thank Larry Greenwood of PNNL for providing the PKA spectrum for 14 MeV neutrons. Also, the authors would like to acknowledge the use of OVITO [21] for visualization.